\documentclass[aps,prl,reprint,showpacs,superscriptaddress]{revtex4-2}
\usepackage{epsfig}
\usepackage[]{natbib}
\usepackage{xspace}
\usepackage{color}
\usepackage{bbold}
\usepackage{tabu}
\usepackage{mathtools}
\usepackage{amsmath}
\usepackage{amssymb}
\usepackage{feynmp-auto}
\usepackage{tikz}

 \usepackage{hyperref}

\hypersetup{
	colorlinks=true,
	linkcolor=blue,
	citecolor=blue,
	urlcolor=blue
}

\newcommand{\Ra}{\langle R \rangle}
\newcommand{\Rb}{\langle R^2 \rangle}
\newcommand{\Rc}{\langle R^3 \rangle}

\newcommand{\ps}[1]{\langle \hat{\psi}(#1) \rangle}

\begin{document}

\title{ Diffusion in Quenched Random Environments: Reviving Laplace's First Law of Errors
}

\author{Lucianno Defaveri}
\affiliation{Department of Physics, Institute of Nanotechnology and Advanced Materials, Bar-Ilan University, Ramat-Gan 52900, Israel}
\author{Eli Barkai}
\affiliation{Department of Physics, Institute of Nanotechnology and Advanced Materials, Bar-Ilan University, Ramat-Gan 52900, Israel}

\begin{abstract}

Laplace's first law of errors, which states that the frequency of an error can be represented as an exponential function of the error magnitude, was overlooked for many decades but was recently shown to describe the statistical behavior of diffusive tracers in disordered, glassy-like media. 
While much is known about this behavior, a key ingredient is still missing: 
the relationship between this observation and diffusion in a quenched 
random environment. We address this problem using the trap model, 
deriving lower and upper bounds on the particle packet for large displacements.
 Our results demonstrate that both bounds exhibit Laplace-like laws. We further establish a connection between the density of energy traps $\rho(E)$, 
 and the observed behavior, showing that the phenomenon is truly universal, albeit with constants that depend on temperature and the level of disorder.

\end{abstract}

\maketitle

{\em Introduction.}
Diffusion in heterogeneous random environments is observed in various scientific domains, from tracers in glasses \cite{Chaudhuri2007,Aberg2021,Denny2003}, to molecules in cellular environments \cite{Barkai2012,Waigh2023} and charge carriers in amorphous solids \cite{Scher1975}. The theoretical exploration of the relationship between quenched disorder and diffusion is extensive \cite{Haus1987,Bouchaud1990,Weiss1996,Douglas1997,Havlin2000}. Classical theories predict a Gaussian spreading packet with a re-normalized diffusion constant in the long-time limit and under weak disorder conditions \cite{Bouchaud1990,Metzler2000,Metzler2014}. However, the emergence of exponential tails in the spreading packet of particles, as widely observed in laboratory and in simulations, revolutionized the paradigms of diffusion phenomenon \cite{Weeks2000,Chaudhuri2007,Hapca2009,Wang2009,Toyota2011, Stuhrmann2012,Wang2012, Guan2014, Stylianidou2014, Soares2014, Lampo2017, Chakraborty2020,Xue2020, Miotto2021, Pastore2021, Rusciano2022, Pastore2022,Rusciano2022,Hu2023,Schramma2023,Yang2024,Burov2024}. In many of these systems, the disorder is not too strong,  leading to normal diffusion characterized by the mean square displacement $\langle x^2 \rangle \propto t$ \cite{Wang2009,Wang2012}.
Stochastic models such as diffusing diffusivity \cite{Chubynsky2014,Jain2017} and Brownian yet non-Gaussian models \cite{Chechkin2017,Postnikov2020} describe this phenomenon and its implications, including, for example, a significant impact on the rate of chemical reactions \cite{Sposini2024}. 

These observations are related to the work of Laplace, who formulated two laws of errors \cite{Laplace1774,Laplace1781,Hamdi2024}. The first law states that the error distribution decays exponentially with the size of the error, while the second law asserts that the error distribution is quadratic with the error size. The first law is deeply connected to recent observations of exponential tails in diffusion processes, while the second law is intimately related to the central limit theorem and the normal distribution.

Chaudhuri, Bertheir, and Kob \cite{Chaudhuri2007} demonstrated how Laplace-like tails appear in continuous-time random walks (CTRW) \cite{Metzler2014}, promoting the idea that exponential Laplace-like tails are indeed universal \cite{Chaudhuri2007,Barkai2020,Wang2020,Pozo2021,Hamdi2024}. However, the CTRW model is a mean-field theory in which the time between jumps are independent identically distributed random variables, and thus, mathematically, the renewal approach can be used. A natural question is whether and under what conditions Laplace-like behavior arises in systems with quenched (i.e., time-independent) disorder \cite{Luo2018,Sokolov2021}. Despite its widespread significance, the precise mechanisms and criteria for this phenomenon in spatially quenched systems remain unresolved.

For particles diffusing in a random environment, disorder appears in various forms, such as energy and entropic traps or barriers. Potential traps of differing depths induce correlations in the motion of the tracer upon re-visitation, a feature shared across other types of disorder.
A major study of diffusion in these systems has focused on the long-time limit of the spreading particle packet. Novel effects, such as the transition from normal to anomalous diffusion, have been well explored \cite{Derrida1983,Bouchaud1990,Bouchaud1990a,Monthus1996,Monthus2003,Bertin2003,Burov2011,Hofling2013,Akimoto2016,Illien2018,Akimoto2019}. Our goal is to study a different limit, relevant to experiments, where the displacement $x$ is large but time $t$ is intermediate. This limit allows us to focus on Laplace's first law. 
In single tracer experiments, a common feature is the hopping of particles among traps.
The average number of hops does not need to be large for the tracer packet's density to exhibit universal features at large displacements. This is because large displacements correspond to many hops, leading to the emergence of universal statistical laws in the tail.
Experimentally, if one waits long enough, packets of particles often transition to a normal distribution, in line with Laplace's second law. 

{\em Trap Model.} 
We will consider a random walk in a random environment on a one-dimensional lattice with unit lattice spacing \cite{Haus1987}.
The probability of finding the particle at position $x$ at time $t$ is $P(x,t)$, where initially
the particle is on the origin, $x=0$.
The evolution of the system is governed by a master equation with typical gain and loss terms:
\begin{equation}
\frac{ \partial P(x,t) }{ \partial t} = - R_x P(x,t) + \frac{ R_{x+1} }{2} P(x+1, t) + \frac{R_{x-1}}{2} P(x-1,t) .
\label{eq01}
\end{equation}
The transition rates $R_x> 0$ are time-independent, mutually independent, 
identically distributed random variables drawn from a common
probability density function (PDF) $f(R)$. 
This model describes a particle moving on a one-dimensional lattice,
 where residence times at each site are exponentially distributed with a rate $R_x$.
The dynamics are non-biased, with the particle hopping to $x\pm 1$ with equal probability $1/2$
after waiting at $x$. 
The key observable is $\langle P(x,t) \rangle$, where the brackets indicate an average
over the quenched disorder $\langle P(x,t) \rangle = \int_{0} ^\infty ... \int_{0}
 ^\infty   \Pi_{y=-\infty} ^{\infty}  f(R_y) {\rm d} R_y P(x, t) $.
This averaging mimics a set of experiments all conducted in specific, mutually independent environments,
which is a common situation in single particle tracking protocols.  

{\em Ordered system.} When all rates are identical and equal $r$, the system is ordered and Eq.\,(\ref{eq01}) gives \cite{Haus1987,Seki2024}:
\begin{equation}
P(x,t) = \exp( - r t) I_{|x|} ( rt )
\label{eq:prob-ordered}
\end{equation}
where $I_{|x|}(\cdot)$ is the modified Bessel function of the first kind.
For large times and $x \propto \sqrt{t}$, the solution
is Gaussian as expected from the central limit theorem. However, we are interested in a different limit, where $x$ is large and $t$ is fixed. Using the asymptotic limit of the modified Bessel function, we find an exponential decay with logarithmic corrections, see supplemental material (SM) \cite{SM}
\begin{equation}
P(x,t) \sim \exp\left[ - |x| \ln \left( { 2 |x| \over e r t } \right) \right]
\frac{ e^{ - r t }}{\sqrt{ 2 \pi |x|}}.
\label{eq02a}
\end{equation}
Rewriting, 
\begin{equation}
\lim_{|x| \to \infty } { - \ln P(x,t)  \over |x| \ln\left( { 2 |x| \over 
 e r t} \right) } = 1
\label{eq02b}
\end{equation}
so the factor $\exp( - r t) /\sqrt{2 \pi |x|}$ in Eq. (\ref{eq02a}), while important for comparison to data, 
contributes insignificantly to the asymptotics.
Such Laplace-like tails of $P(x,t)$ are found in full generality for CTRW models \cite{Chaudhuri2007,Luo2018,Barkai2020,Wang2020,Sokolov2021,Pozo2021,Hamdi2024,Luo2024}.

\begin{figure}[b]
    \centering
    \includegraphics[width=0.85\linewidth]{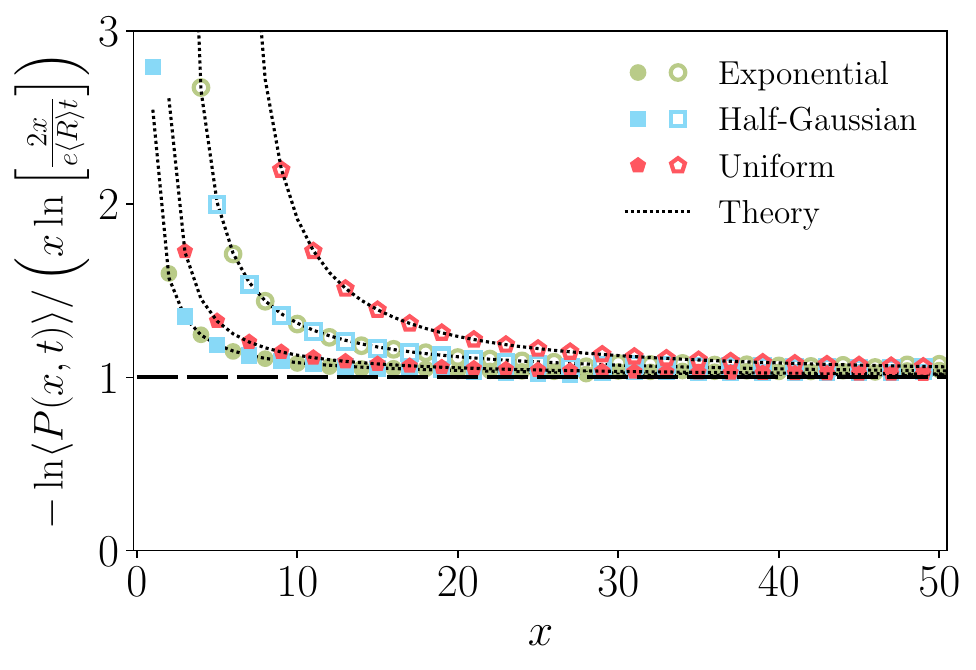}
    \caption{The log of the disorder-averaged probability $\langle P(x,t) \rangle$, obtained through numerical simulations, divided by $x\ln[(2x)/(e \langle R \rangle t)]$, illustrating the validity of Eq.\,(\ref{eq:asymp-general}), and hence the Laplace-like behavior. This universal law is shown for different types of disorder $\rho(E)$ (colored symbols shown in the caption) for $t=1$ (solid symbols), and $t=5$ (hollow symbols). The details of the different types of disorder, listed in the caption, can be found in Table\,\ref{tab:energy_densities} and in the SM. 
    The meaningful correction due to the backtracking paths (black-dotted line), found in Eq.\,(\ref{eq:lower-bound-theory}),  shows excellent agreement already for intermediate values of $x$.}
    \label{fig:px-asymp}
\end{figure}

To relate this result to the properties of the underlying paths, we highlight the direct path, which starts on $x=0$, jumps to $1$, and sequentially until reaching $x$. We denote this path with an arrow $\rightarrow$. Thus,  $\mbox{Prob}(\rightarrow)$ is the probability of
observing such a path in the laboratory. 
In the SM we obtain this probability exactly for any $x$ and $t$, showing that $P(x,t) = \mbox{Prob}(\rightarrow)$ for large $|x|$, implying that the mechanism of finding the particles at large distances is dominated by direct paths.
We investigate  whether similar trends and relations
extend to disordered systems.
Specifically, we derive general lower and upper bounds for $\langle P(x,t)\rangle$,
demonstrating  that the Laplace tails of $\langle P(x,t) \rangle$
 are universal, being insensitive
 to the type of disorder. We find the scales and constants involved,
showing their dependence on the model details. 
Additionally,  we demonstrate the effectiveness of the lower bound and 
present a systematic method for its improvement.
While the first-order bound, obtained using the direct path principle, 
is useful for illustrating Laplace tails, higher-order bounds highlight the significance of correlations between trajectories and disorder. Thus our goal is threefold: firstly to show the universality of Laplace-like tails, secondly to show how backtracking paths also yield exponential decay of the density, and thirdly, to find the impact of disorder on this basic phenomenon.

{\em The density of states.}
Our analysis focuses on generic distributions of $f(R)$, suitable for thermal
and non-thermal settings \cite{Woillez2020}, provided that the rate moments 
$\langle R \rangle$ and $\langle R^2 \rangle$ are finite.
For systems coupled to a heat bath with temperature $T$,
 the escape rates from the traps are described by
Arrhenius law \cite{Hangi1990,Kumar2024}
 $R_x = r \
\exp( - E_x/T)$, where $E_x>0$ is the energetic depth of the trap
and $r$ is a microscopical rate constant, with the
 Boltzmann constant set to unity.
  The depth of traps $E_x$ are independent
random variables drawn from the density of states $\rho(E)$ \cite{Bouchaud1990,Denny2003}. While our theory generally holds, the different disorder models used are shown in Table\,\ref{tab:energy_densities}. One of the basic open questions is: what is the relation between the density of states $\rho(E)$ and Laplace-like diffusion? In glassy systems common choices for $\rho(E)$ are the exponential \cite{Monthus1996,Scher1975,Shafir2024} and the Gaussian \cite{Dyre1987,Bassler1987,Diezemann2011,Scalliet2021}.
\begin{table}[h!]
\centering
\begin{tabular}{|c|c|c|}
\hline
\textbf{Name} & $\rho(E)$ ($E>0$) & $\langle R^q \rangle$ \\
\hline
Exponential & ${\frac{1}{k_B T_g} e^{-\frac{E}{k_B T_g}} } $ & $\frac{T}{T + q T_g}$ \\
\hline
Gaussian & $\sqrt{\frac{2}{\pi \sigma_E^2}}e^{-\frac{E^2}{2 \sigma_E^2}}$ & $e^{\frac{\sigma_E^2 q^2}{2 (k_B T)^2}} \text{erfc}\left(\frac{\sigma_E \, q}{\sqrt{2} k_B T }\right)$ \\
\hline
Uniform & $\frac{1}{E_\mathrm{max}}$ \,, \, $E < E_\mathrm{max}$ & $\frac{1 - e^{-E_\mathrm{max} q/k_B T}}{E_\mathrm{max} q / k_B T}$ \\
\hline
\end{tabular}
\caption{Energy densities $\rho(E)$ and corresponding expressions for $\langle R^q \rangle$ for different distributions.}
\label{tab:energy_densities}
\end{table}

{\em Lower bound.} Consider a particle starting at $0$ and
found at $x$ at time $t$. The probability $P(x,t)$ accounts 
for all possible paths between these points. We again employ $\mbox{Prob}(\rightarrow)$, the probability of obtaining the direct path in the disordered system, and clearly, for any $x$ and\,$t$
\begin{equation}
P(x,t) > \mbox{Prob}(\rightarrow).
\label{eq04}
\end{equation}
In a directed path, the particle remains for random times
$\{\tau_0,\tau_1, ... \}$ at traps numbered $0,1, \cdots$.
We introduce the indicator function
$\mathbb{I} \left( \sum_{i=0} ^{x-1} \tau_i < t < \sum_{i=0} ^{x} \tau_i \right)$,
 which is $1$ if the condition holds true and $0$ otherwise. For simplicity, we will assume $x>0$.
Averaging over all the waiting times (WT), which are nonidentical
exponentially distributed random variables, we have
\begin{equation} 
\mbox{Prob}(\rightarrow) =\left(  {1 \over 2} \right)^x \left\langle 
\mathbb{I} \left( \sum_{i=0} ^{x-1} \tau_i < t < \sum_{i=0} ^{x} \tau_{i+1}  \right)\right \rangle_{{\rm WT}}.
\label{eq04a}
\end{equation}
This expression
depends on the rates $\{ R_x\}$ describing the specific system under study and discussed further in the SM.  
To correctly weigh the directed path, 
we use the fact that
after waiting, the probability of jumping to the right is $1/2$, and hence the weight of a directed path is $(1/2)^x$, as given in Eq. (\ref{eq04a}).

\begin{figure}
    \centering
    \includegraphics[width=0.85\linewidth]{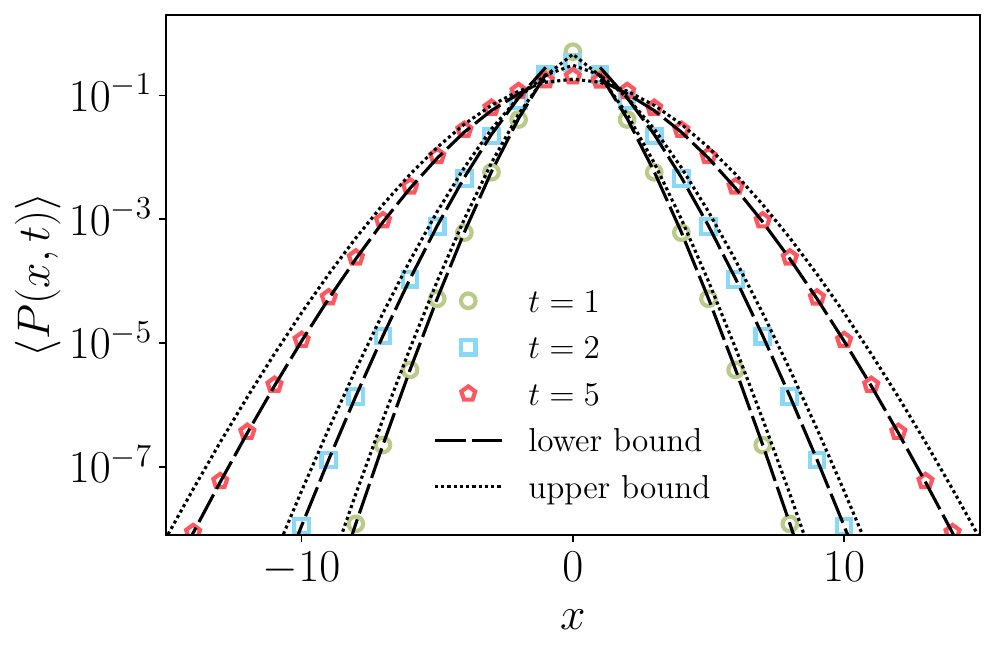}
    \caption{The disorder-averaged probability $\langle P(x,t) \rangle$ for different times, shown in the legend. The lower bound, Eq.\,(\ref{eq08}) (dashed black lines), and the upper bound, Eq.\,(\ref{eq02a}) (dotted black lines), are plotted. In the large $x$ limit, we see that $\langle P(x,t) \rangle$ shows an exponential-like decay, in accordance with Laplace law. We have used an exponential type of disorder, for $\rho(E)$, see Table\,\ref{tab:energy_densities}, with $T/T_g = 6$.}
    \label{fig:bounds}
\end{figure}

 We now average 
Eqs. (\ref{eq04},\ref{eq04a})
over the disorder: 
\begin{equation}
\langle P(x,t) \rangle > 
\langle \mbox{Prob}(\rightarrow) \rangle \, .
\label{eq05}
\end{equation}
This expression is determined by the disorder-averaged waiting time PDF denoted $\langle \psi(\tau) \rangle$ \cite{Scher1975,Bouchaud1990,Metzler2000}. 
The process under study is described using the master equation (\ref{eq01}). Hence, the
 time spent on each trap is exponentially distributed.
%
It follows that $\langle \psi(\tau) \rangle = \int_{0} ^\infty R\exp( - R \tau) f(R) {\rm d} R$
%
which in terms of the density of states is
\begin{equation}
\langle \psi(\tau) \rangle 
= 
-{  {\rm d} \over \rm d \tau } \int_{0} ^\infty \exp\left[ -  \exp\left( - E/T \right) r \tau \right] \rho(E) {\rm d } E.
\label{eq06a}
\end{equation}
The key property of the directed path is that it does not retract, so
the waiting times that the particle samples are 
non-correlated.
Using Eqs. (\ref{eq04a},\ref{eq05})
and Laplace transforms we find the first-order
lower bound. 
Let $\langle \hat{ \psi}(s) \rangle= \int_{0} ^\infty \exp( -s \tau ) \langle \psi(\tau) \rangle {\rm d} \tau$ be the Laplace transform of
$\psi(\tau)$ and similarly $\widehat{\mbox{Prob}}(\rightarrow)$ the Laplace transform of $\mbox{Prob} (\rightarrow)$.
We arrive at our first main result (see  SM) 
\begin{equation}
 \langle \widehat{\mbox{Prob}}(\rightarrow) \rangle= \left( {1 \over 2} \right)^{x}
 { 1 - \langle \hat{\psi}(s) \rangle \over s} \left[\langle  \hat{\psi}(s) \rangle \right]^x.
\label{eq07}
\end{equation}
Inverting this formula to the time domain in different limits is a classical problem and can be performed numerically. Using the saddle point approximation (see SM), we have for large $x$
\begin{equation}
\langle  \mbox{Prob} (\rightarrow) \rangle \sim
\exp\left[ - |x| \ln \left({ 2 |x| \over e \langle R \rangle t} \right) \right]
{ e^{ - {\langle R^2 \rangle t \over \langle R \rangle}}  \over 
\sqrt{ 2 \pi |x|} } \, .
\label{eq08}
\end{equation}
Eq. (\ref{eq08})
implies that only the first two moments of the random
 rates are needed, given by:
\begin{equation}
\langle R^q \rangle =  r^q  \int_{0} ^\infty \exp( - q E / T) \rho(E) {\rm d} E,
\label{eq09}
\end{equation}
with $q=1,2$. 
The Laplace transform of the density of states, given by Eq. (\ref{eq09}), determines the moments and hence the large $x$ limit of the density of spreading particles.  
Importantly, from Eq. (\ref{eq05},\ref{eq08}), for any type of disorder $\rho(E)$ and any temperature, the packet decays no faster than exponentially with increasing distance $x$, with logarithmic corrections that are difficult to detect experimentally. We find that Eq.\,(\ref{eq02b}), true for ordered systems, can be extended, using Eq.\,(\ref{eq08}), to any type of disorder to
\begin{equation}
    \lim_{|x| \to \infty } { - \ln \langle P(x,t) \rangle  \over |x| \ln\left( { 2 |x| \over 
 e \langle R \rangle t} \right) } \geq 1 \, . \label{eq:asymp-general}
\end{equation}
We argue that the bound in the previous equation is actually an identity, as we expect that the direct path for large $x$ is the dominating path. A strong indication for this is that the correction terms we calculate below, stemming from back tracking paths, are negligible when $x \to \infty$.
Further numerical evidence is provided in Fig.\,\ref{fig:px-asymp} for different types of disorder. We see that Eq.\,(\ref{eq:asymp-general}) is consistent with Laplace's law.

\begin{figure}
    \centering
    \includegraphics[width=\linewidth]{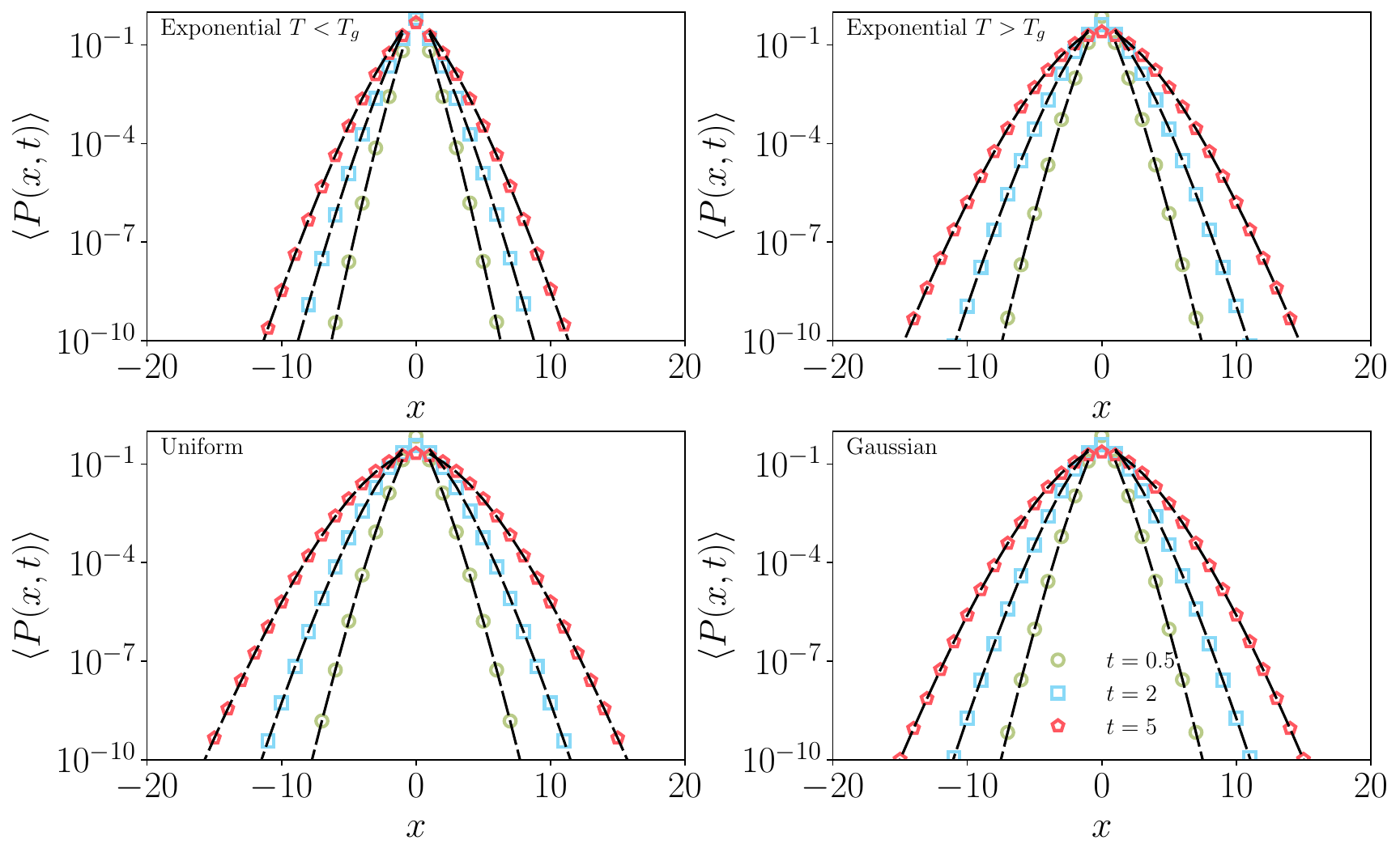}
    \caption{The disorder-averaged probability $\langle P(x,t) \rangle$ for different times, shown in the legend for four different disorder types, labeled on top of each panel. The theoretical lower bound in Eq.\,(\ref{eq:lower-bound-theory}) is plotted for comparison. The theoretical prediction is very accurate regardless of the type of disorder. This is true even for the exponential case with $T<T_g$, where the diffusion, at long times, is anomalous as the mean squared displacement behaves as $\langle x^2\rangle \sim t^{T/T_g}$. The bound in Eq.\,(\ref{eq:lower-bound-theory}) features the Laplace-like exponential decay, with pre-exponential corrections, a phenomenon which is universal and valid for any density of states $\rho(E)$. The pre-exponential terms are crucial when the data is plotted for not too large $x$, as is done in experiments.}
    \label{fig:pxt-models}
\end{figure}


To demonstrate that Laplace-like tails are universal, we derive an upper bound for $\langle P(x,t) \rangle$. In the infinite temperature limit, which is equivalent to the ordered system, a particle diffuses to larger distances faster compared to a finite temperature system because the waiting time at each trap is statistically short. At this limit, disorder is irrelevant since the escape rate from each trap is fixed and denoted by $r$. Therefore, for large $x$, $\langle P(x,t) \rangle < P_{\text{ordered}}(x,t)$, where $P_{\text{ordered}}$ is defined in Eq.\,(\ref{eq:prob-ordered}). It follows that Eq.\,(\ref{eq02a}) is an upper bound that decays exponentially. Given that both upper and lower bounds exhibit Laplace tails, the system, regardless of temperature or density of states, presents Laplace-like tails for sufficiently large $x$. In Fig.\,\ref{fig:bounds}, we plot both lower and upper bounds, showing that the lower bound always saturates, while the upper bound becomes exact for high temperatures (as shown in the SM).

{\em Single-turn path}. The convergence of $\langle P(x,t) \rangle$ to $\langle \mathrm{Prob}(\to) \rangle$
can be very slow for not-so-large values of $x$, and hence, we will now calculate corrections due to indirect paths. 
Consider the path starting on lattice point 0 and ending on $x$, $\{ 0,1,2,1,2,3,4,5,\ldots x \}$. This is a path with a single reversal, where the sites $1$ and $2$  are visited twice. More generally, the backtracking can take place on any site of the lattice point connecting 0 and $x$, see details in SM. The following diagram represents a single-turn path, where the path moves to the right, makes a loop, and then continues to the right:
$\begin{array}{c}
            \includegraphics[scale=0.75]{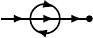}
            \end{array}$.
These paths can be used to improve our lower bound and show that the exponential decay is not merely a feature of direct paths. Here,  the correlations between the trajectory and the disordered environment are crucial, unlike the direct path scenario.  Specifically, when resampling traps, the rate of escape from the trap at the first visit is the same as that in the second.
Note that we have two sites visited twice, except when the reversal takes place at the start or end of the path (see details in the SM). The total number of single-turn paths is $x+2$, with the number of paths where the loop is not on the edges is $x-1$. These paths represent the largest contribution, allowing us to ignore the contribution of the edge paths. A complete description of all possible paths can be found in the SM. A bound that works well for large $x$ is then
\begin{equation}
\langle P (x,t) \rangle > \left\langle \mbox{Prob} \left(\rightarrow\right ) \right\rangle + (x-1) \left\langle
\mbox{Prob} \left(
\begin{array}{c}
            \includegraphics[scale=0.75]{zig_1.pdf}
            \end{array}
\right) \right\rangle \, .
\label{eq11}
\end{equation} 
A calculation in the SM yields the Laplace transform:
\begin{equation}
\left\langle \widehat{\mbox{Prob}}\left(
\begin{array}{c}
            \includegraphics[scale=0.75]{zig_1.pdf}
            \end{array}
\right) \right\rangle =  \frac{\langle \hat{\psi}(s)^2 \rangle^2 \langle \hat{\psi}(s) \rangle^{|x|-2}}{2^{|x|+2}}  \left( \frac{1 - \langle \hat\psi(s) \rangle}{s} \right) 
\label{eq12}
\end{equation} 
Since in the backtracking paths, the particle will revisit some sites, we see that the previous formula depends on the second moment of the disorder-averaged waiting time $\langle \hat \psi (s)^2 \rangle = \langle R^2/(s + R)^2 \rangle$. 
Neglecting correlations, which is in general invalid, these terms can be replaced by $\langle \hat \psi (s)^2 \rangle = \langle \hat \psi (s) \rangle^2$.
The Laplace transform in Eq.\,(\ref{eq12}) can be inverted numerically or using a saddle point method, valid for large $x$. In the latter case, we find in the SM 
\begin{equation}
\left\langle \mbox{Prob}\left(
\begin{array}{c}
            \includegraphics[scale=0.75]{zig_1.pdf}
            \end{array}
\right) \right\rangle \approx \frac{e^{-\frac{\langle R^2 \rangle}{\langle R \rangle}t}}{\sqrt{2 \pi |x|}} \left[ \frac{e \langle R \rangle t}{2 |x|} \right]^{|x|} \, \left[ \frac{\Rb t}{2 \Ra |x|} \right]^2 \, . \label{eq:one-zig-correction}
\end{equation} 
Eqs.\,(\ref{eq11},\ref{eq:one-zig-correction}) show that the contributions from single turn paths are by order $1/x$ smaller if compared with the contribution of the directed path. Hence, to find an explicit formula for the bound Eq.\,(\ref{eq11}), we need to correct Eq.\,(\ref{eq08}) to the same order. This task is performed in the SM, where we find our main result
\begin{widetext}
\begin{equation}
        \langle P(x,t) \rangle \gtrsim \frac{e^{-\frac{\langle R^2 \rangle}{\langle R \rangle}t}}{\sqrt{2 \pi |x|}} \exp\left[ - |x| \ln \left( { 2 |x| \over e \langle R \rangle t } \right) \right] \,  \left\{ 1 - \frac{1}{12 |x|} + \frac{t}{|x|} \left[ \frac{\langle R^2 \rangle}{\langle R \rangle} - \langle R \rangle \right] + \frac{t^2}{|x|} \left[ \frac{\langle R^3 \rangle}{\langle R \rangle} -\frac{3}{4} \frac{\langle R^2 \rangle^2}{\langle R \rangle^2} \right] \right\} \, . \label{eq:lower-bound-theory}
\end{equation}
\end{widetext}
As we can clearly see from the previous equation, the Laplace law is observed even for the first corrections. The validity of Eq.\,(\ref{eq:lower-bound-theory}) is shown in Figs.\,\ref{fig:px-asymp} and \,\ref{fig:pxt-models}. The figures clearly demonstrate the effectiveness of Eq.\,(\ref{eq:lower-bound-theory}) for the estimation of the probability packet,  for large, though finite, $x$.

Our theoretical research, based on a master equation describing hopping in a disordered system, supports the appearance of exponential-like tails of $\langle P(x,t) \rangle$, as observed in many systems. The insensitivity of our main results to details such as the density of trap states or temperature suggests a feature that transcends specific models. To summarize, we find that for large $x$, the disorder-averaged probability $\langle P(x,t) \rangle$ behaves in accordance with the Laplace law of errors, that is, it decays exponentially (with logarithm corrections). This striking result is universal, i.e., independent of the specific nature of the disorder. Importantly, the backtracking corrections presented in Eq.\,(\ref{eq:lower-bound-theory}) provide crucial $1/x$ corrections. These corrections capture the correlations induced by the quenched disorder. Further, for not-so-large values of $x$, these corrections are important when comparing theory and simulations.

\noindent {\bf Acknowledgements:}
The support of Israel Science Foundation's grant 1614/21 is acknowledged.


%

\clearpage

\setcounter{equation}{0}
\setcounter{figure}{0}
\renewcommand{\theequation}{SM\arabic{equation}}
\renewcommand{\thefigure}{S\arabic{figure}}

\onecolumngrid

\section{Supplementary Material}

\subsection{I.~~~Ordered system}

In this section, we derive Eqs. 
(\ref{eq:prob-ordered},
\ref{eq02a}), an explicit formula for $\mbox{Prob}(\rightarrow)$ and the asymptotic
relation $P(x,t) \sim \mbox{Prob}(\rightarrow)$ in the absence of disorder. 

We solve the master equation
for the probability of finding the walker at $x$ at time $t$
\begin{equation} 
\frac{\partial P (x,t)}{\partial t}= - r P(x,t) + {r \over 2}\left[  P(x,t) + P(x,t) \right] \, ,
\label{eqR01}
\end{equation}
where the initial condition is $P(0,0)=1$ and zero otherwise, 
indicating the particle starts at $x=0$. 
The transition rate $r$ is uniform over all lattice points $x$,
so, from symmetry, $P(x,t)= P(-x,t)$.   
Clearly, Eq.\,(\ref{eq01}) and Eq.\,(\ref{eqR01}) are the same when $R_x=r$.
Taking the Fourier transform of $P(x,t)$, denoted by ${\tilde P} (k, t)$,
we obtain $\dot{\tilde{P} } (k,t) = - (1 - \cos k )  \tilde{P}(k,t)$.
Applying the inverse Fourier transform 
\begin{equation}
P(x,t) = { 1 \over 2 \pi} \int_{-\pi} ^\pi \exp\left[ - i k x - r t( 1 - \cos k) \right] {\rm d} k .
\label{eqR02}
\end{equation}
Evaluating the integral, we get
$P(x,t) = \exp( - r t) I_{|x|} ( r t )$, Eq. 
(\ref{eq:prob-ordered}) in the main text.

The asymptotic expansion of the modified Bessel function 
 for large orders $\nu$  and fixed real arguments $z\neq 0$
is given by \cite{NIST} 
\begin{equation}
I_\nu (z) \sim {1 \over \sqrt{ 2 \pi \nu} } \left( { e z \over 2 \nu} \right)^\nu.
\label{eqR03}
\end{equation}
where $e=2.718\ldots $  is the base of the natural logarithm. 
Using Eq.\,(\ref{eq:prob-ordered}) in the main text, for large $|x|$ and fixed $t$,  we find:
\begin{equation}
P(x,t) \sim \exp\left( - r t \right) { 1 \over \sqrt{2 \pi |x|} } 
\left( { e r  t \over 2 |x|} \right)^{ |x|} 
\label{eqR03}
\end{equation}
or equivalently Eq.
(\ref{eq02a})
 in the main text. 
Reformulating this limit, we have the Laplace-like tails
\begin{equation}
\lim_{|x| \to \infty} { - \ln P(x,t) \over |x| \ln\left( \frac{ 2 |x|} {e r t} 
\right)} = 1.
\label{eqR05}
\end{equation}
This demonstrates the asymptotic behavior of $P(x,t)$ for large $|x|$. 

This result is related to directed
paths, symbolized with $\rightarrow$. 
In the underlying stochastic process, let $Q(n;t)$ be the number
of jumps $n$  made in the interval $(0,t)$. This satisfies the master equation
$\dot{Q}(n;t) = - r Q (n;t) + r Q (n-1; t)$, which yields
the Poisson 
distribution 
\begin{equation}
Q(n;t) = { (r t)^n \exp( - r t) \over n!}.
\label{eqR06}
\end{equation}
The probability of finding a direct path $\rightarrow$,
starting on $0$ and ending at $x>0$, namely a path that jumps 
through
$0,1,2,3 ... x$, at random times,  is  
the probability of observing $|x|$ jumps, all in the same direction:
\begin{equation}
\mbox{Prob}(\rightarrow) = { (r t)^{|x|}  \exp\left( - r t \right) \over |x|!} \left( {1 \over 2}\right)^{|x|} .
\label{eqR07}
\end{equation}
We used the model's assumption that the
 probability of jumping left or right after waiting on any lattice
point is $1/2$. 
Using the Stirling approximation 
$|x|!\sim \sqrt{ 2 \pi |x|} ( |x|/e)^{|x|}$ for large $|x|$,
we obtain the Laplace-like  asymptotics
\begin{equation}
\mbox{Prob}(\rightarrow) 
 \sim \exp\left[ - |x| \ln \left( { 2 |x| \over e r t } \right) \right]
{ e^{ - r t } \over \sqrt{ 2 \pi |x|} } 
\label{eqR07}
\end{equation}
or
\begin{equation}
\lim_{|x| \to \infty} { - \ln\mbox{Prob}(\rightarrow) \over |x| \ln\left( \frac{ 2 |x|} {e r t} 
\right)} = 1.
\label{eqR08}
\end{equation}
This matches the result for $P(x,t)$ Eqs. (\ref{eq02a},\ref{eqR05}), 
leading to
the asymptotic relation
 $P(x,t) = \mbox{Prob} (\rightarrow)$ for large
$x$, as mentioned in the main text.

\subsection{II.~~~Disordered system: direct path approach}

In this section, we evaluate the Laplace transform 
 of the probability of finding a directed path at position \( x \) at time \( t \). 
More specifically, considering a disordered system, 
Eq. (\ref{eq04a}) reads
\begin{equation} 
\mbox{Prob} (\rightarrow) =\left(  {1 \over 2} \right)^x \left\langle 
\mathbb{I} \left( \sum_{i=0} ^{x-1} \tau_i < t < \sum_{i=0} ^{x} \tau_{i+1}  \right)\right \rangle_{{\rm WT}}.
\label{eq04aaaa}
\end{equation}
Our goal is to further clarify this formula, 
and subsequently find the disorder-averaged $\langle \mbox{Prob} (\rightarrow)\rangle$
Eq. 
(\ref{eq07}). We recall that the subscript WT implies an average over the waiting times (WT) in each trap, defined in the main text and also below more precisely. 
Further, we use the pulse function $\mathbb{I}\left( \sum_{i=0} ^{x-1} \tau_i < t < \sum_{i=0} ^{x} \tau_i \right)$, which is one if the restriction is obeyed and zero otherwise, as mentioned in the main text.

First, we consider the underlying process in a fixed random environment. Consider the direct path starting at \(0\). Let the trapping times in the states along this path be \(\tau_0, \tau_1, \cdots\). These random trapping times are constrained by the measurement time:
\begin{equation}
t = \tau_0 + \tau_1 + \cdots + \tau_x^{*}.
\label{eqSM01}
\end{equation}
Since the particle is at the \(x\) trap at time \(t\), the dwelling time \(\tau_x^{*}\) at \(x\) does not end with a jump; instead, it ends when the external clock reaches \(t\). In related problems, this time is called the backward recurrence time.

It is useful to define the jumping moments when the particle hops between one trap to its nearest neighbors \cite{Wang2020}. The process starts at \(t=0\), and jumps take place at times \( t_1=\tau_0\), \(t_2=\tau_0 + \tau_1\), etc. In particular \(t_{x-1}=\sum_{a=0}^{x-1} \tau_{a}< t\) and hence \(\tau_x^{*} = t - t_x\). We will also use the transition event taking place at time \(t_x = \sum_{a=0}^{x} \tau_{a}\) which is clearly larger than \(t\).
Using Eq. (\ref{eq01}) \(\tau_a\), are independent non-identically distributed exponential random variables, whose PDFs are
\begin{equation}
\psi_a (\tau_a) = R_a \exp( - R_a \tau_a)
\label{eqSM03}
\end{equation}
and we will later use the Laplace transform
\begin{equation}
\hat{\psi}_s(s) = \int_{0} ^\infty  \psi_a (\tau_a ) \exp( - s \tau_a) \, \mathrm{d } \tau_a
= \frac{R_a}{R_a + s}.
\label{eq:waiting-time-dist-laplace}
\end{equation}
Rewriting Eq. (\ref{eq04aaaa}) in terms of jump moments
\begin{equation}
\mbox{Prob} (\rightarrow) = \left( {1 \over 2} \right)^x \langle \mathbb{I} (t_{x-1} < t < t_x) \rangle_{{\rm WT}}.
\label{eqSM05}
\end{equation}
Note \(t_0 = 0\). The brackets denote an average over the waiting times  \(\tau_0, \tau_1, \cdots\), namely,
\begin{equation}
\mbox{Prob} (\rightarrow)  = \left( {1 \over 2} \right)^x \left[ \int_0^{\infty} R_0 \exp(-R_0 \tau_0) \, \mathrm{d} \tau_0 \cdots \int_0^{\infty} R_x \exp(-R_x \tau_x) \, \mathrm{d} \tau_x \right] \mathbb{I}(t_{x-1} < t < t_x).
\label{eqSM06}
\end{equation}
We now study the Laplace transform \(t \to s\) of this expression, denoted $\widehat{\mbox{Prob}} ( \rightarrow)$. Using
\begin{equation}
\int_{0}^{\infty} \exp(-st) \mathbb{I}(t_{x-1} < t < t_x) \, \mathrm{d} t = \frac{\exp(-st_{x-1}) - \exp(-st_x)}{s}
\label{PeqSM07}
\end{equation}
and the definitions of \(t_x\) and \(t_{x-1}\), we find
\begin{equation}
\widehat{\mbox{Prob}}\left( \rightarrow \right) = \left( {1 \over 2} \right)^x \left\langle \frac{\exp\left[-s(\tau_0 + \tau_1 + \cdots + \tau_{x-1})\right] - \exp\left[-s(\tau_0 + \tau_1 + \cdots + \tau_x)\right]}{s} \right\rangle_{{\rm WT}}
\label{eqSM08}
\end{equation}
The waiting times are independent random variables, and the averaging gives
\begin{equation}
\widehat{\mbox{Prob}}\left( \rightarrow \right) = \left( { 1 \over 2 } \right)^x
 \frac{\left(\prod_{a=0}^{x-1} \frac{R_a}{R_a + s} - \prod_{a=0}^{x} \frac{R_a}{R_a + s}\right)}{s}, 
\label{eq:direct-path-laplace-1}
\end{equation}
where we used Eq. (\ref{eq:waiting-time-dist-laplace}).
As mentioned in the text this expression is sensitive to the set of rates
in the system. 

 We now average over disorder. We define $\langle \hat{\psi}(s) \rangle = \langle R_a / (R_a + s)\rangle$ and
 exploit the fact that
the rates
$\{ R_a \}$ are independent identically distributed random variables with a common PDF $f(R)$
\begin{equation}
\langle \hat{\psi} (s) \rangle = \int_{0} ^\infty f(R) { R \over R + s} {\rm d} R. 
\label{eq:avg-psi-s}
\end{equation}
For the trap model, where rates are given by Arrhenius law, the maximum of the rate is the microscopical rate $r$,
so $f(R) = 0$ for $R>r$ and the upper limit of the integration
is $r$. 
Using Eq. (\ref{eq:direct-path-laplace-1}) we get Eq. (\ref{eq07}) in the main text. 
Clearly $\langle \hat{\psi}(s) \rangle$ is the Laplace transform of the disordered average distribution
of waiting times
$\langle \psi(\tau) \rangle$, Eq. (\ref{eq06a}) in the letter.

\subsection{III.~~~Direct path probability through saddle point approximation}

Our goal in this section is to use the saddle point approximation to obtain a theoretical expression for the probability of the direct path as an asymptotic series of $1/x$. The Laplace transform of the probability of the direct path, given in Eq.\,(\ref{eq:direct-path-laplace-1}), can be written as
\begin{align}
    \widehat{\mathrm{Prob}}(\to) &= \frac{1}{2^x} \left( \frac{1 - \hat{\psi}_x(s)}{s} \right) \prod_{a=0}^{x-1} \hat{\psi}_a (s)  \, , \label{eq:general-prob-laplace} 
\end{align}
where we make use of the Laplace transform of the waiting time distribution defined in Eq.\,(\ref{eq:waiting-time-dist-laplace}). The inverse Laplace transform of Eq.\,(\ref{eq:general-prob-laplace}) yields the probability of the direct path at time $t$.  Clearly, Eq.\,(\ref{eq:general-prob-laplace}), depends on the $x+1$ rates, which, as mentioned, are themselves random variables. Therefore, since the direct path does not revisit any site, we can write the disorder average of the product as
    \begin{align}
        \langle \widehat{\mathrm{Prob}}(\to)  \rangle &= \frac{1}{2^x}\left\langle \left( \frac{1 - \hat{\psi}_x(s)}{s} \right) \prod_{a=0}^{x-1} \hat{\psi}_a (s) \right\rangle =   \frac{1}{2^x} \left( \frac{1 - \ps{s}}{s} \right)  \ps{s}^x \, . \label{eq:general-prob-laplace-average}
    \end{align}
    To calculate the inverse of the Laplace transform, we will make use of the Mellin integral, that is,
    \begin{align}
            \langle \mathrm{Prob}(\to) \rangle &= \int_{0^+ -i \infty}^{0^+ +i \infty} \frac{ds}{2\pi i} \frac{e^{st}}{2^x} \left( \frac{1 - \ps{s}}{s} \right)  \ps{s}^x  \, . \label{eq:inverse-Laplace-before-g}
    \end{align}
    In the large $x$ limit, we can perform the integral in the previous equation using the method of the steepest descent or the saddle point approximation \cite{Daniels1954}. This method consists of deforming the integration path in the complex plane so that it crosses a saddle point $s^*$, which is in the real line. As is well known, in this new path, the largest contribution to the integral comes from the vicinity of $s^*$.

    We re-write Eq.\,(\ref{eq:inverse-Laplace-before-g}) as
    \begin{align}
            \langle \mathrm{Prob}(\to) \rangle &= \frac{1}{2^x} \int_{s^* -i \infty}^{s^* + +i \infty} \frac{ds}{2\pi i} e^{g(s)}  \, , \label{eq:g(s)-int-def}
    \end{align}
    where the integration is performed on the new path and the auxiliary function $g(s)$ is defined, explicitly, as
    \begin{align}
        g(s) \equiv s t + x \ln \ps{s} + \ln \left[1 - \ps{s} \right] - \ln s \, .  \label{eq:g(s)-direct-def}
    \end{align}
    The saddle point must satisfy $g'(s^*) = 0$. Since the largest contribution to the integral comes from the vicinity of $s^*$, we can expand the auxiliary function $g(s)$ as
    \begin{align}
        g(s) \approx g(s^*) + \frac{g^{(2)}(s^*)}{2!}(s-s^*)^2 + \frac{g^{(3)}(s^*)}{3!}(s-s^*)^3 + \frac{g^{(4)}(s^*)}{4!}(s-s^*)^4 \, .
    \end{align}
    Here we have included terms up to the fourth-order since we will calculate corrections to the leading term. These terms are important for Eq.(\ref{eq:lower-bound-theory}) in the main text but can be neglected for Eq.(\ref{eq:asymp-general}).
    Plugging the expansion in the previous equation into Eq.\,(\ref{eq:g(s)-int-def}), and performing the change of variable $i z = (s-s^*)$ we write
    \begin{align}
        \langle \mathrm{Prob}(\to) \rangle = \int_{s^* - i \infty}^{s^* + i \infty} \frac{ds}{2 \pi i} e^{g(s)} &\approx e^{g(s^*)} \int_{s^* - i \infty}^{s^* + i \infty} \frac{ds}{2 \pi i} e^{\frac{g^{(2)}(s^*)}{2} (s - s^*)^2 + \frac{g^{(3)}(s^*)}{6} (s - s^*)^3 + \frac{g^{(4)}(s^*)}{24} (s - s^*)^4 } \\
        &\approx e^{g(s^*)} \int_{-\infty}^{\infty} \frac{dz}{2 \pi} e^{-\frac{g^{(2)}(s^*)}{2} z^2 + \frac{g^{(3)}(s^*)}{6} \, i z^3 + \frac{g^{(4)}(s^*)}{24} z^4 } \, .
    \end{align}
    As usual, the integration path was selected to ensure that $g^{(2)}(s^*) > 0$. In the next subsections, we will show the derivatives at the critical point scale with $x$, up to the leading order, as $g^{(n)}(s^*) \propto x^{1-n}$. This allows us to expand the $z^3$ and $z^4$ terms in the exponent, and perform the Gaussian integral, that is,
    \begin{align}
       \langle \mathrm{Prob}(\to) \rangle &\approx e^{g(s^*)} \int_{-\infty}^{\infty} \frac{dz}{2 \pi} e^{-\frac{g^{(2)}(s^*)}{2} z^2}\left\{ 1+ \frac{g^{(3)}(s^*)}{6} \, i z^3 + \frac{g^{(4)}(s^*)}{24} z^4 - \frac{g^{(3)}(s^*)^2}{72} z^6 \right\} \nonumber \\
        &\approx \frac{e^{g(s^*)} }{\sqrt{2 \pi g^{(2)}(s^*)}} \left\{ 1 + \frac{g^{(4)}(s^*)}{8 g^{(2)}(s^*)^2} - \frac{5 g^{(3)}(s^*)^2}{24 g^{(2)}(s^*)^3} \right\} \, , \label{eq:prob-theo-general-expression}
    \end{align}
    where the $i z^3$ term is null due to symmetry. For large $x$, $g^{(4)}(s^*)/g^{(2)}(s^*)^2 \propto g^{(3)}(s^*)^2/g^{(2)}(s^*)^3 \propto 1/x$, and all extra terms to the expansion are, at least, of order $O(1/x^2)$. 
    
    Now we obtain an analytical expression for $\langle \mathrm{Prob}(\to) \rangle$ by solving Eq.\,(\ref{eq:inverse-Laplace-before-g}) in the asymptotic limit of large $x$, considering that the waiting times are exponentially distributed and that the energy density is $\rho(E)$. The saddle point $s^*$ can be found by finding the critical point of $g(s)$, which is defined in Eq.\,(\ref{eq:g(s)-direct-def}), that is
    \begin{align}
        g'(s^*) = 0 \to t + x \frac{\langle \hat \psi ' (s^*) \rangle}{\langle \hat \psi (s^*) \rangle} - \frac{\langle \hat \psi ' (s^*) \rangle}{1 - \langle \hat \psi(s^*) \rangle } - \frac{1}{s^*} = 0 \, . \label{eq:g(s)-prime-null}
    \end{align}
    In the large $x$ limit, we have the critical $s^* \propto x$. Therefore, we are also interested in the large $s$ limit of $\ps{s}$. From Eq.\,(\ref{eq:avg-psi-s}), in the limit of large $s$, we have
    \begin{align}
        \ps{s} &= \int_0^\infty f(R) \frac{R}{R+s} dR \approx \int_0^\infty f(R) \left( \frac{R}{s} - \frac{R^2}{s^2} + \frac{R^3}{s^3} + O(1/s^4) \right) dR \nonumber \\
        &\approx \frac{\Ra}{s} - \frac{\Rb}{s^2} + \frac{\Rc}{s^3} \, .
    \end{align}
    We plug the expression for $\ps{s}$ in the previous equation into Eq.\,(\ref{eq:g(s)-prime-null}) to find
    \begin{align}
       0 &= t - \frac{x+1}{s^*} + \frac{\langle R \rangle + x \frac{\langle R^2 \rangle}{\langle R \rangle}}{{s^*}^2} + \frac{\langle R^2 \rangle - 2 \langle R^2 \rangle - \frac{x (2 \langle R \rangle \langle R^2 \rangle - \langle R^2 \rangle^2 )}{\langle R \rangle^2}}{{s^*}^3} + O({s^*}^4) \, . \label{eq:g-prime-asym-s}
    \end{align}
    The saddle point can be expanded in a series of $x$, as
    \begin{align}
        s^* &\approx s_1 x + s_0 + \frac{s_{-1}}{x} + O(1/x^2) \, ,
    \end{align}
    where the values of $s_1$, $s_0$ and $s_{-1}$ can be found by plugging the previous equation into Eq.\,(\ref{eq:g-prime-asym-s}), to obtain an expression for $g'(s^*)$ as a series of $1/x$,
    \begin{align}
        g'(s^*) &\approx \left[ t + \frac{1}{s_0} \right] - \frac{1}{x}\left[ \frac{s_1}{s_0^2}  - \frac{\Rb}{\Ra s_0^2} + \frac{1}{s_0} \right] + O(1/x^2) \, = 0 . \label{eq:series-s0-s1}
    \end{align}
    The values of $s_1$, $s_0$, and $s_{-1}$ are obtained by imposing that every order of $1/x$ in the previous equation be null independently. We will show next that, up to $1/x$ terms, $g(s^*)$ does not depend on $s_{-1}$.

    The values of $s_1$ and $s_0$ are obtained from Eq.\,(\ref{eq:series-s0-s1}) as
    \begin{align}
        s_1 = \frac{1}{t} \, \, \, , \, \, \,  s_0 = \frac{1}{t} - \frac{\Rb}{\Ra} \, , \label{eq:s0-s1}
    \end{align}
    leading to the expression of the saddle point
    \begin{align}
        s^* = \frac{x}{t} + \left[ \frac{1}{t} - \frac{\Rb}{\Ra} \right] + O(1/x) \, .
    \end{align}
    From Eq.\,(\ref{eq:g-prime-asym-s}), we find that the leading term of the $n$-th derivative of $g(s)$ at the saddle point is
    \begin{align}
        g^{(n \geq 2)}(s^*) \approx \frac{(-1)^n (n-1)!}{x^{1-n}}\, ,
    \end{align}
    and the correction term in Eq.\,(\ref{eq:prob-theo-general-expression}) of higher order derivatives is
    \begin{align}
        \frac{g^{(4)}(s^*)}{8 g^{(2)}(s^*)^2} - \frac{5 g^{(3)}(s^*)^2}{24 g^{(2)}(s^*)^3} = \frac{\frac{6}{x^3}}{\frac{8}{x^2}} - \frac{5 \frac{16}{x^4}}{24 \frac{1}{x^3}} = - \frac{1}{12x} \, .
    \end{align}
    Finally, we write the leading term of Eq.\,(\ref{eq:prob-theo-general-expression}) as
    \begin{align}
        \frac{e^{g(s^*)} }{\sqrt{2 \pi g^{(2)}(s^*)}} = \frac{e^{g(s^*) - \frac{1}{2} \ln g^{(2)}(s^*)} }{\sqrt{2 \pi}} \, .
    \end{align}
    We will now expand the terms in the exponent to order $1/x$. The second term in the previous equations exponent is straightforward to expand, we have $\ln g^{(2)}(s^*) \approx - 2 \ln (x/t) + \ln x - \frac{1}{x}$. The first term expansion is
    \begin{align}
        g(s^*) &\approx  - (x+1) \ln (s_1 x) +  x \left[ s_1 t + \ln \Ra  \right] + \left[s_0 t -  \frac{s_0}{s_1} - \frac{\Rb}{\Ra s_0} \right] \nonumber \\
        &+ \frac{1}{x}\left[ \frac{s_0^2}{2 s_1^2} + s_{-1} \left( t - \frac{1}{s_1} \right) + \frac{\Rb s_0}{\Ra s_1^2} + \frac{\Rc}{\Ra s_1^2} - \frac{\Rb^2}{2 \Ra^2 s_1^2} - \frac{s_0}{s_1} - \Ra \right] \, ,\label{eq:expansion-gstar-s0-s1}
    \end{align}
    which is $s_{-1}$ independent.
    Note that, in previous equation, the only term dependent on $s_{-1}$ is multiplied by $t - 1/s_0$, which is null when we replace the value of $s_0$ found in Eq.\,(\ref{eq:s0-s1}). Therefore, despite that in a general case we require the $s_{-1}/x$ term to expand $g(s^*)$ to order $1/x$, in our particular function it is not necessary.

    Replacing $s_0$ and $s_1$ from Eq.\,(\ref{eq:s0-s1}) into Eq.\,(\ref{eq:expansion-gstar-s0-s1}) to find
    \begin{align}
        g(s^*) \approx  - (x+1) \ln \left(\frac{x}{t}\right) +  x \left[ 1 + \ln \Ra  \right] - \left[\frac{\Rb}{\Ra}t \right] + \frac{t^2}{x} \left[ \frac{\Rc}{\Ra} - \frac{\Rb^2}{\Ra^2} \right]  + \frac{t}{x}  \left[ \frac{\Rb - \Ra^2}{\Ra} \right] - \frac{1}{2x} \, ,
    \end{align}
    and finally, 
    \begin{align}
        \langle \mathrm{Prob}(\to) \rangle &\approx \frac{e^{- \frac{\Rb}{\Ra}t}}{\sqrt{2 \pi x}} \left( \frac{e \Ra t}{2 x} \right)^x  \left\{ 1 - \frac{1}{12x} + \frac{t}{x}  \left[ \frac{\Rb - \Ra^2}{\Ra} \right\} + \frac{t^2}{x} \left[ \frac{\Rc}{\Ra} - \frac{\Rb^2}{\Ra^2} \right] \right] \, . \label{eq:prob-direct-full}
    \end{align}
    For convenience, we highlight the leading order as
    \begin{align}
        \langle \mathrm{Prob}(\to) \rangle_0 = \frac{e^{- \frac{\Rb}{\Ra}t}}{\sqrt{2 \pi x}} \left( \frac{e \Ra t}{2 x} \right)^x \, , \label{eq:leading-prob-direct}
    \end{align}
    with the first correction being
    \begin{align}
        \langle \mathrm{Prob}(\to) \rangle_1 = \langle \mathrm{Prob}(\to) \rangle_0 \left\{  \frac{t^2}{x} \left[ \frac{\Rc}{\Ra} - \frac{\Rb^2}{\Ra^2} \right]  + \frac{t}{x}  \left[ \frac{\Rb - \Ra^2}{\Ra} \right] - \frac{1}{12x} \right\} . \label{eq:correction-prob-direct-leading}
    \end{align}

    \begin{figure}
        \centering
        \includegraphics[width=0.8\linewidth]{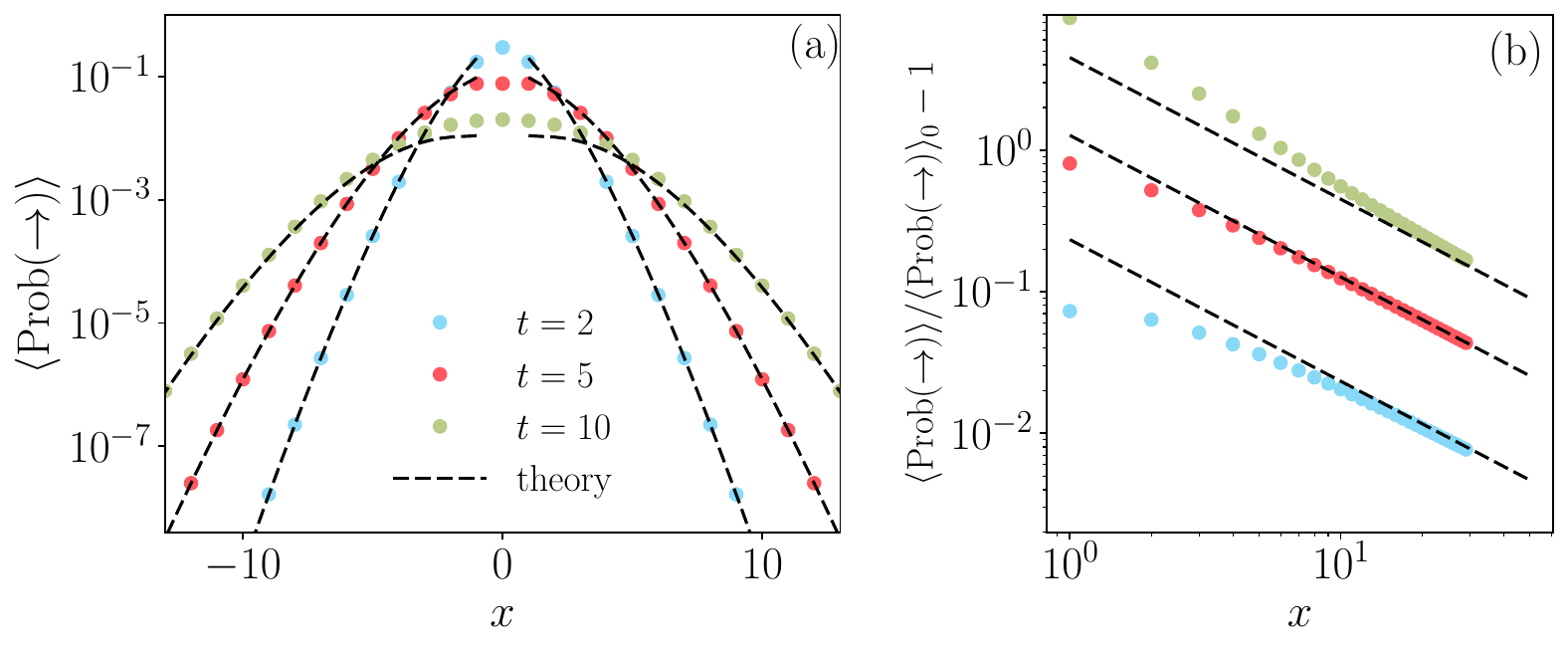}
        \caption{In panel (a): The average over disorder of the probability of the direct path $\langle \mathrm{Prob} (\to) \rangle$, obtained by numerically performing the inverse Laplace transform of Eq.\,(\ref{eq:general-prob-laplace-average}) versus the position $x$ for different times (shown in legend of panel (a), common for all panels). The dashed black line in panel (a) is equivalent to our asymptotic theoretical expansion for large $x$ in Eq.\,(\ref{eq:prob-direct-full}). In panel (b): we show the relative difference between the numerical $\langle \mathrm{Prob} (\to) \rangle$ and the leading approximation $\langle \mathrm{Prob}(\to) \rangle_0$, Eq.\,(\ref{eq:leading-prob-direct}). The dashed black line is the first order correction in Eq.\,(\ref{eq:correction-prob-direct-leading}) divided by the first order correction in Eq.\,(\ref{eq:leading-prob-direct}). We have used $T=2 T_g$.}
        \label{fig:direct_2}
    \end{figure}

    \subsection{IV.~~~Lower bound probability including the contribution of single-turn paths}

    We have used the direct path to find a lower bound for the probability of spreading particles $\langle P(x,t) \rangle$. For not-so-large values of $x$, the probability is no longer clearly dominated by the direct path, and now possesses a meaningful contribution from paths that meander and zig-zag before reaching the final position. We now improve this bound, including what we term zig-zag paths. 

\begin{figure}
        \centering
        \includegraphics[width=0.9\linewidth]{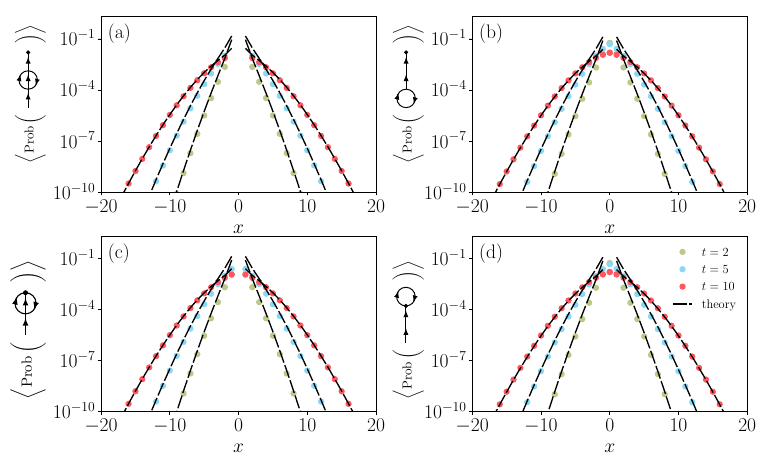}
        \caption{On the panels (a), (b), (c) and (d) we have the numerical probabilities Eqs.\,(\ref{eq:laplace-zig-1}), (\ref{eq:laplace-zig-2}), (\ref{eq:laplace-zig-3}) and (\ref{eq:laplace-zig-4}) respectively (symbols) for different times shown in the legend of panel (d), common for all panels. The dashed black lines represent the theoretical large $x$ approximation in Eqs.\,(\ref{eq:zig-theo-1}), (\ref{eq:zig-theo-2}), (\ref{eq:zig-theo-3}) and (\ref{eq:zig-theo-4}). We have used $T=2T_g$. }
        \label{fig:enter-label}
    \end{figure}

    The total number of paths, given that the particle performed $n$ jumps, is given, of course, by Newton's Binomial $\binom{n}{{[n+x]}/{2}}$. The next corrections come from paths with $n=x+2$ jumps, and there are $x+2$ possible combinations of these paths. Here, we show the four possible types of path containing one zig-zag:
    \begin{itemize}
        \item[(a)] One zig-zag in the middle ($x-1$ possible paths):
        \begin{align}
            \textsc{Path} &= 0 \to 1  \to \cdots \to a \to a+1 \to a \to a+1 \to \cdots \to x \equiv \begin{array}{c}
            \includegraphics[scale=0.75]{zig_1.pdf}
            \end{array} \label{eq:zig-middle} \, \, .
        \end{align}
        \item[(b)] One zig at the start (1 possible path):
        \begin{align}
            \textsc{Path} &=  0 \to -1 \to 0 \to 1 \to 2 \to \cdots \to x \equiv  \begin{array}{c}
            \includegraphics[scale=0.75]{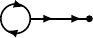}
            \end{array} \, \, .
        \end{align}
        \item[(c)] One back zig at the end (1 possible path):
        \begin{align}
            \textsc{Path} &= 0 \to 1 \to 2  \to \cdots \to x \to x-1 \to x \equiv  \begin{array}{c}
            \includegraphics[scale=0.75]{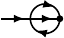}
            \end{array} \, \, .
        \end{align}
        \item[(d)] One forward zig at the end (1 possible path):
        \begin{align}
            \textsc{Path} &= 0 \to 1 \to 2  \to \cdots \to x \to x+1 \to x \equiv  \begin{array}{c}
            \includegraphics[scale=0.75]{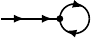}
            \end{array} \, \, .
        \end{align}
    \end{itemize}
    The assigned diagrams are a useful short-cut notation for paths.
    The expression for the Laplace transform of the probability, shown in Eq.\,(\ref{eq:general-prob-laplace}), can be immediately extended to the zig-zag paths. We take, for example, the path outlined in Eq.\,(\ref{eq:zig-middle}). In that case, the Laplace transform of a single realization of disorder is
    \begin{align}
        \widehat{\mathrm{Prob}}\left( \begin{array}{c}
            \includegraphics[scale=0.75]{zig_1.pdf}
            \end{array} \right) = \hat{\psi}_0(s) \hat{\psi}_1(s) \cdots \hat{\psi}_a(s) \hat{\psi}_{a+1}(s) \hat{\psi}_a(s) \hat{\psi}_{a+1}(s) \cdots \hat{\psi}_{x-1}(s) \left( \frac{1 - \hat{\psi}_x(s)}{s} \right) \, . \label{eq:example-zig-zag-path}
    \end{align}
    As we mentioned before, the distribution of waiting times of a given site $\psi_a(t) = R_a \, e^{- R_a t}$ depends on the local rate $R_a$, which in turn is a random variable. Unlike the direct path, the particles will now revisit previous sites, in the example in Eq.\,(\ref{eq:example-zig-zag-path}), sites $a$ and $a+1$ are revisited. Therefore, when we take the ensemble average, their contribution to the total Laplace transform is $\langle \hat\psi(s)^2 \rangle \geq \langle \hat\psi(s) \rangle^2$, with the equality holding for the ordered case. The large $s$ limit, explicitly, is
    \begin{align}
        \langle \hat\psi(s)^2 \rangle \approx \frac{\Rb}{s^2} - \frac{2 \Rc}{s^3} \, .
    \end{align}
    We extend this to all possible paths to write the expression for the Laplace transforms of the probabilities as
    \begin{align}
            \left\langle \widehat{\mathrm{Prob}}\left( \begin{array}{c}
            \includegraphics[scale=0.75]{zig_1.pdf}
            \end{array} \right) \right\rangle &= \frac{1}{2^{x+2}} \langle \hat{\psi}(s)^2 \rangle^2 \langle \hat{\psi}(s) \rangle^{x-2} \left( \frac{1 - \langle \hat\psi(s) \rangle}{s} \right) \, , \label{eq:laplace-zig-1} \\
            \left\langle \widehat{\mathrm{Prob}} \left( \begin{array}{c}
            \includegraphics[scale=0.75]{zig_2.pdf}
            \end{array} \right) \right\rangle  &= \frac{1}{2^{x+2}} \langle \hat{\psi}(s)^2 \rangle \langle \hat{\psi}(s) \rangle^{x} \left( \frac{1 - \langle \hat\psi(s) \rangle}{s} \right) \, , \label{eq:laplace-zig-2}
            \\
            \left\langle \widehat{\mathrm{Prob}} \left( \begin{array}{c}
            \includegraphics[scale=0.75]{zig_3.pdf}
            \end{array} \right) \right\rangle &= \frac{1}{2^{x+2}} \langle \hat{\psi}(s)^2 \rangle \langle \hat{\psi}(s) \rangle^{x-1} \left( \frac{\langle \hat{\psi}(s) \rangle - \langle \hat{\psi}(s)^2 \rangle}{s} \right) \, , \label{eq:laplace-zig-3} \\
            \left\langle \widehat{\mathrm{Prob}} \left( \begin{array}{c}
            \includegraphics[scale=0.75]{zig_4.pdf}
            \end{array} \right) \right\rangle &= \frac{1}{2^{x+2}} \langle \hat{\psi}(s) \rangle^{x+1} \left( \frac{\langle \hat{\psi}(s) \rangle - \langle \hat{\psi}(s)^2 \rangle}{s} \right) \, . \label{eq:laplace-zig-4}
    \end{align}
    We calculate the Laplace inverse through the same method as we used for the direct path in the previous section. First, we re-write,
    \begin{align}
            \left\langle \mathrm{Prob}\left( \begin{array}{c}
            \includegraphics[scale=0.75]{zig_1.pdf}
            \end{array} \right) \right\rangle &= \frac{1}{2^{x+2}} \int_{0^+ -i \infty}^{0^+ +i \infty} \frac{ds}{2\pi i} e^{g_1(s)} \, , \\
             \left\langle \mathrm{Prob}\left( \begin{array}{c}
            \includegraphics[scale=0.75]{zig_2.pdf}
            \end{array} \right) \right\rangle &=  \frac{1}{2^{x+2}} \int_{0^+ -i \infty}^{0^+ +i \infty} \frac{ds}{2\pi i} e^{g_2(s)} \, ,
            \\
            \left\langle \mathrm{Prob}\left( \begin{array}{c}
            \includegraphics[scale=0.75]{zig_3.pdf}
            \end{array} \right) \right\rangle &= \frac{1}{2^{x+2}} \int_{0^+ -i \infty}^{0^+ +i \infty} \frac{ds}{2\pi i} e^{g_3(s)} \, , \\
            \left\langle \mathrm{Prob}\left( \begin{array}{c}
            \includegraphics[scale=0.75]{zig_4.pdf}
            \end{array} \right) \right\rangle &=  \frac{1}{2^{x+2}} \int_{0^+ -i \infty}^{0^+ +i \infty} \frac{ds}{2\pi i} e^{g_4(s)} \, ,
    \end{align}
    with the auxiliary functions being
    \begin{align}
        g_1(s) &\equiv s t + (x-2) \ln \langle \hat{\psi}(s) \rangle + 2 \ln \langle \hat{\psi}(s)^2 \rangle + \ln \left[ 1 - \langle \hat{\psi}(s) \rangle \right] - \ln s \, , \\
        g_2(s) &\equiv s t + x \ln \langle \hat{\psi}(s) \rangle + \ln \langle \hat{\psi}(s)^2 \rangle + \ln \left[ 1 - \langle \hat{\psi}(s) \rangle \right] - \ln s \, , \\
        g_3(s) &\equiv  s t + (x-1) \ln \langle \hat{\psi}(s) \rangle + \ln \langle \hat{\psi}(s)^2 \rangle + \ln \left[ \langle \hat{\psi}(s) \rangle - \langle \hat{\psi}(s)^2 \rangle \right] - \ln s \, , \\
        g_4(s) &\equiv s t + (x+1) \ln \langle \hat{\psi}(s) \rangle  + \ln \left[ \langle \hat{\psi}(s) \rangle - \langle \hat{\psi}(s)^2 \rangle \right] - \ln s \, .
    \end{align}
    We expand the saddle point for the different paths as an asymptotic series for large $x$ as $s^* \approx s_1 x + s_0$. The saddle points for the different paths are, for orders higher than $1/x$, identical. We can find the analytical expression through
    \begin{align}
        g_i'(s^*) \approx \left[t - \frac{1}{s_1} \right] + \frac{1}{x} \left[ \frac{s_0}{s_1^2} + \frac{\Rb}{\Ra s_1^2} - \frac{3}{s_1}  \right] + O(1/x^2) = 0 \, ,
    \end{align}
    and therefore,
    \begin{align}
        s_1 = \frac{1}{t} \, \, , \, \, s_0 = \frac{3}{t} - \frac{\Rb}{\Ra} \, .
    \end{align}
    This allows us to write the value of the auxiliary functions $g_i$'s at the saddle point, up to order $1/x$, as
    \begin{align}
        g_1(s^*) &\approx  - (x+1) \ln \left( \frac{x}{t} \right) + x[1 + \ln \Ra] - \left[ \frac{\Rb}{\Ra}t \right] + 2 \ln \left( \frac{\Rb t}{\Ra x} \right) \\
        g_2(s^*) &\approx  - (x+1) \ln \left( \frac{x}{t} \right) + x[1 + \ln \Ra] - \left[ \frac{\Rb}{\Ra}t \right] + \ln \left( \frac{\Rb t^2}{x^2} \right) \\
        g_3(s^*) &\approx  - (x+1) \ln \left( \frac{x}{t} \right) + x[1 + \ln \Ra] - \left[ \frac{\Rb}{\Ra}t \right] + \ln \left( \frac{\Rb t^2}{x^2} \right) \\
        g_4(s^*) &\approx  - (x+1) \ln \left( \frac{x}{t} \right) + x[1 + \ln \Ra] - \left[ \frac{\Rb}{\Ra}t \right] + 2 \ln \left( \frac{\Ra t}{x} \right) \, .
    \end{align}
    The probability contribution from each zig-zag path is calculated using
    \begin{align}
        \frac{1}{2^{x+2}} \int_{0^+ -i \infty}^{0^+ +i \infty} \frac{ds}{2\pi i} e^{g_i(s)} \approx \frac{1}{2^{x+2}} \sqrt{ \frac{2\pi}{|g_i''(s^*)|} } \, e^{g_i(s^*)} \, ,
    \end{align}
    where $g_i''(s^*) = t^2/x$. The complete asymptotic expressions for the probabilities are
    \begin{align}
            \left\langle \mathrm{Prob}\left( \begin{array}{c}
            \includegraphics[scale=0.75]{zig_1.pdf}
            \end{array} \right) \right\rangle & \approx  \frac{e^{-\frac{\langle R^2 \rangle}{\langle R \rangle}t}}{\sqrt{2 \pi x}} \left[ \frac{e \langle R \rangle t}{2 x} \right]^x \, \left[ \frac{\Rb t}{2 \Ra x^2} \right]^2 + O(1/x^3) \, , \label{eq:zig-theo-1} \\
             \left\langle \mathrm{Prob}\left( \begin{array}{c}
            \includegraphics[scale=0.75]{zig_2.pdf}
            \end{array} \right) \right\rangle & \approx \frac{e^{-\frac{\langle R^2 \rangle}{\langle R \rangle}t}}{\sqrt{2 \pi x}} \left[ \frac{e \langle R \rangle t}{2 x} \right]^x \, \frac{\Rb t^2}{4 x^2} + O(1/x^3) \, ,
            \label{eq:zig-theo-2}  \\
            \left\langle \mathrm{Prob}\left( \begin{array}{c}
            \includegraphics[scale=0.75]{zig_3.pdf}
            \end{array} \right) \right\rangle & \approx  \frac{e^{-\frac{\langle R^2 \rangle}{\langle R \rangle}t}}{\sqrt{2 \pi x}} \left[ \frac{e \langle R \rangle t}{2 x} \right]^x \, \frac{\langle R^2 \rangle t^2}{4 x^2} + O(1/x^3) \, , \label{eq:zig-theo-3} \\
            \left\langle \mathrm{Prob}\left( \begin{array}{c}
            \includegraphics[scale=0.75]{zig_4.pdf}
            \end{array} \right) \right\rangle & \approx  \frac{e^{-\frac{\langle R^2 \rangle}{\langle R \rangle}t}}{\sqrt{2 \pi x}} \left[ \frac{e \langle R \rangle t}{2 x} \right]^x \, \left[ \frac{\Ra t}{2 x} \right]^2 + O(1/x^3) \, \label{eq:zig-theo-4}  .
    \end{align} 
    From the previous set of equations, we see that the probability contribution from the zig-zag paths is of order $1/x^2$ smaller than the leading contribution from the direct path. The largest contribution, for large $x$, comes from the first path, zig-zag in the middle of the trajectory $\begin{array}{c}
            \includegraphics[scale=0.75]{zig_1.pdf}
            \end{array}$. This is due to the fact that there are $x-1$ possible such paths, leading to a complete contribution of order $1/x$, which is the same order of the first correction to the direct path, see Eq.\,(\ref{eq:prob-direct-full}). Using Eqs.\,(\ref{eq:prob-direct-full}) and (\ref{eq:zig-theo-1}), our asymptotic theoretical prediction for the lower bound becomes
    \begin{align}
        \langle P(x,t) \rangle & \gtrsim \langle \mathrm{Prob} (\to) \rangle + (x-1) \left\langle \mathrm{Prob}\left( \begin{array}{c}
            \includegraphics[scale=0.75]{zig_1.pdf}
            \end{array} \right) \right\rangle \nonumber \\
            & \gtrsim \frac{e^{-\frac{\langle R^2 \rangle}{\langle R \rangle}t}}{\sqrt{2 \pi x}} \left[ \frac{e \langle R \rangle t}{2 x} \right]^x \,  \left\{ 1 - \frac{1}{12x} + \frac{t}{x} \left[ \frac{\langle R^2 \rangle}{\langle R \rangle} - \langle R \rangle \right] + \frac{t^2}{x} \left[ \frac{\langle R^3 \rangle}{\langle R \rangle} -\frac{3}{4} \frac{\langle R^2 \rangle^2}{\langle R \rangle^2} \right] \right\} \, . \label{eq:lower-bound-theory-app}
    \end{align}
    This gives us Eq.\,(\ref{eq:lower-bound-theory}) in the main text.

\section{Details on figures}
The numerical results are obtained by solving the system of differential equations in Eq.\,(\ref{eq01}), with reflecting boundary conditions, using a Runge-Kutta scheme. The values of $R_x$ are chosen at the start and remain fixed for the duration of the run, generating the random rates in accordance with Table \ref{tab:energy_densities}. After performing $\mathcal{N} = 100000$ such runs, the probability is extracted by taking the average of each run. This yields $\langle P(x,t) \rangle$, here we choose the size of the system to be large enough so that the boundary conditions do not affect the results.

\begin{itemize}
    \item[Fig.\,\ref{fig:px-asymp}:] For the exponential disorder, we used $T = T_g/2$, which leads to $\langle R \rangle = 2/3$. For the Gaussian, we used $\sigma_E = 2 k_B T$, with the first moment $\langle R \rangle \approx 0.336$. For the uniform, we used $E_\mathrm{max} = k_B T/2$, and $\langle R \rangle \approx 0.787$.
    \item[Fig.\,\ref{fig:bounds}:] The model used was the exponential, with $T = 6T_g$. See Table \ref{tab:energy_densities} from the main text.
    \item[Fig.\,\ref{fig:pxt-models}:] For the exponential model with $T<T_g$, we used $T=T_g/2$. The values of the first three moments of the rates are $\langle R \rangle = 2/3$, $\langle R^2 \rangle = 1/2$, and $\langle R^3 \rangle = 2/5$. For the second exponential model, now with $T>T_g$, we used $T=2T_g$. The value of the first three moments $\langle R \rangle = 1/3$, $\langle R^2 \rangle = 1/5$, and $\langle R^3 \rangle = 1/7$. For the Gaussian model, we used $\sigma_E=k_B T/2$. The value of the first three moments $\langle R \rangle \approx 0.699$, $\langle R^2 \rangle \approx 0.523$, and $\langle R^3 \rangle \approx 0.412$. For the uniform model, we used $E_\mathrm{max} = k_B T/2$. The value of the first three moments $\langle R \rangle \approx 0.787$, $\langle R^2 \rangle \approx 0.632$, and $\langle R^3 \rangle \approx 0.518$.
\end{itemize}

In all cases, the spacing between lattices is set to unity, also $k_B T = 1$ and $r=1$ ($R_x = r e^{- E_x/k_B T}$). These parameters are entered in the local rates.

\end{document}